\documentclass[PRE,twocolumn,showpacs]{revtex4-1}
\usepackage{amsmath,amssymb,latexsym,epsfig,graphics,epsf}
\usepackage{fixltx2e} 
\usepackage{graphics}
\usepackage{float}
\usepackage{graphicx}
\usepackage{caption}
\usepackage{multirow}
\usepackage{subfigure}

\newcommand{\be}{\begin{equation}}
\newcommand{\ee}{\end{equation}}
\begin{document}
\title{$d^0$-$d$ half-Heusler alloys: A potential class of advanced spintronic materials}
\author{A.\ Dehghan}
\author{S.\ Davatolhagh}
\altaffiliation{Corresponding Author: davatolhagh@shirazu.ac.ir}
\affiliation{Department of Physics, College of Sciences, Shiraz
University, Shiraz 71946, Iran}
\date{\today}
\begin{abstract}
The possibility of ferromagnetic half-metallicity in lithium-based half-Heusler alloys, such as MnPLi, has been theoretically explored before [Damewood et\,al., Phys. Rev. B {\bf 91}, 064409 (2015)], although at optimized lattice constants such lithiated manganese pnictides are predicted to be ordinary ferromagnets and therefore not suitable for spintronic applications. In the following, however, it is shown by first principles density functional calculations that half-Heusler alloys formed by $3d$ transition metals and $d^0$ alkali or alkaline-earth metals, which are defined by the valence electronic configuration $ns^{1,2},(n-1)d^0$, can produce all kinds of half-metallic behavior at optimized lattice constants including the elusive Dirac half-metallicity that is theoretically predicted for the first time in a {\em three-dimensional} prototype MnPK. Together with the predicted magnetic and mechanical stability, this could pave the way for massless and dissipationless spintronics of the future. Among other technologically important features of the prototype Dirac half-metal MnPK, are the maximal moment $d$-shell ferromagnetism, high Fermi velocity of Dirac fermions that is  comparable with that in graphene, and a wide half-metallic gap. Furthermore, by considering the prototype conventional half-metal VSbSr, the introduction of $d^0$ metals is shown to substantially reduce the hull distance and produce {\em true metastability} in the otherwise instable chemical structure of zinc-blende transition metal pnictides--in this case zinc-blende VSb--without altering the `$p$-$d$ exchange' that is mainly responsible for their half-metallicity, thus bringing their realization and potential applications in spintronics a step closer to reality.
\end{abstract}

\maketitle


\section{\label{sec:Introduction}introduction}

The half-metallic (HM) materials with characteristic 100\% spin polarization at the Fermi level, have been regarded as the ideal materials for spintronics \cite{wolf,piket}. Following the seminal work of de Groot et\,al.\ predicting the HM property in half-Heusler (HH) alloy MnNiSb \cite{FirstHF}, much theoretical work has been devoted to finding other HM materials of Heusler family \cite{GD}; transition metal (TM) oxides \cite{Ox1,Ox}; zinc-blende TM pnictides or chalcogenides \cite{Xie,GM03,Liu10}; and non-TM-based $sp$-electron HM ferromagnets \cite{Geshi,Sieberer}. More recently, spin gapless semiconductor (SGS) is introduced as another class of HM materials characterized by an open semiconductor band gap for one spin direction and nearly closed gap for the other \cite{Wang08}. The SGS materials predicted and fabricated so far are mainly the TM-based full-Heusler and inverse full-Heusler compounds with  nearly zero but indirect band gaps \cite{Ouardi,Skaf13,Bainsla15,Wangetal16}. To the best of our knowledge, there has been no report of SGS in stoichiometric half-Heusler compounds (see, also, Ref.\ \cite{Wangetal16}). A particularly interesting kind of SGS that has been shown to exist in a model two-dimensional (2D) ferrimagnetic system is the Dirac half-metal (DHM) \cite{Ishizuka}, also referred to as Dirac SGS \cite{Wang16}, with its characteristic Dirac point linear dispersion for one spin channel--as in graphene \cite{Graphene}--and an open semiconductor band gap for the other. The experimental realization of DHMs is envisioned to pave the way for development of ultra-fast and power-efficient spintronics of the future \cite{Wang16}. Although there have been a number of theoretical proposals for the realization of 2D and quasi-2D DHM compounds \cite{Pardo09,Wang13,Zhang15,Li15,Cai15,Wei16,Gmitra16,Chen16,He16}, the complexity of the structures and/or low magnetic transition temperatures, have rendered their fabrication and possible device application a formidable challenge \cite{Wang16}. It is therefore imperative to find potential three-dimensional (3D) DHM compounds characterized by a greater prospect for realization and room temperature applications. To this end, we theoretically propose, among others, the prototype $d^0$-$d$ Dirac half-metal HH-MnPK.

The spin degeneracy of electronic bands $E_n(\vec{k},\uparrow )=E_n(\vec{k},\downarrow)$, originates from the simultaneous effect of time-reversal and spatial-inversion symmetry. As pointed out by de Groot et\,al.~\cite{FirstHF}, the lack of spatial-inversion symmetry in HH structure, in addition to the broken time-reversal symmetry as in ordinary ferromagnets, further removes the spin degeneracy of electronic bands and  may result in robust half-metallicity. This makes the HH $C1_b$ structure, space group F$\bar{4}$3m (No.\ 216 in the international tables of crystallography), particularly interesting from both theoretical and technological point of view \cite{Xiao10,Nat1,Nat2,Casper12}. More explicitly, the crystal structure of a ternary HH compound consists of a fcc Bravais lattice with three atom basis situated along the cube diagonal at (0,0,0), ($\frac{1}{4}$,$\frac{1}{4}$,$\frac{1}{4}$), and ($\frac{1}{2}$,$\frac{1}{2}$,$\frac{1}{2}$) Wyckoff position. It is worth noting that the zinc-blende (ZB) structure is rather similar to the HH structure, except that the ($\frac{1}{2}$,$\frac{1}{2}$,$\frac{1}{2}$) Wyckoff position is left out as a void. In the $d^0$-$d$ HH alloys hereby introduced, the $3d$ TM atom takes the (0,0,0) position, the main body $sp$ atom occupies the {\em unique} ($\frac{1}{4}$,$\frac{1}{4}$,$\frac{1}{4}$) middle site, and the $d^0$ metal atom defined by the valence electronic configuration $ns^{1,2}$, $(n-1)d^0$ is situated at the ($\frac{1}{2}$,$\frac{1}{2}$,$\frac{1}{2}$) Wyckoff position. By definition, therefore, the $d^0$ metals include the alkali metals except Li and Na, and  the alkaline-earth metals except Be and Mg. The above arrangement with the electronegative $sp$ atom positioned in between the electropositive metallic atoms is also referred to as the  $\beta$-phase in the literature, and is found to be energetically most favorable, together with a ferromagnetic alignment of the TM atoms (see, Fig.~\ref{Fig3}). The properties of $d^0$-$d$ HH alloys in their metastable $\gamma$-phase with the TM atom at the unique middle site have been addressed elsewhere \cite{DD}. It must also be pointed out that lithiated manganese pnictides, such as MnPLi, have been investigated before in the half-Heusler $\beta$-phase for possible signs of half-metallic ferromagnetism \cite{Damewood}. Such lithium-based half-Heusler alloys although mechanically stable as indicated by the phonon dispersion curves, are found to be ordinary metallic ferromagnets at their optimized lattice constants, and therefore not suitable for spintronic applications. It must, however, be noted that Li is not a $d^0$ alkali metal.

In the following, the electronic structure results at optimized lattice constants, the structural stability tests, and Curie temperatures are reported for the prototype $d^0$-$d$ HH alloys in the $\beta$-phase. The details of procedures and computational methods are given in Sec.\ \ref{sec:comp}. In Sec.\ \ref{sec:VSbSr} by considering VSbSr as a prototype conventional half-metal, it is shown that the introduction of $d^0$ metals drastically reduces the distance from the convex hull and generates true metastability as revealed by phonon band structure calculations in the  otherwise instable chemical structure of half-metallic zinc-blende TM pnictides--in this case ZB-VSb--thus increasing the likelihood of their experimental realization and potential spintronic applications. It is a matter of considerable interest to find in Sec.\ \ref{sec:MnPK} that another prototype $d^0$-$d$ HH alloy, MnPK, exhibits Dirac half-metallicity, a property that is being reported for the first time in 3D structures. The paper is concluded with a summary and final remarks in Sec.\ \ref{sec:Summary}.

\section{\label{sec:comp}Computational methods}
The electronic structure calculations are performed on the basis of spin-polarized density functional theory (DFT) as implemented by {\scriptsize WIEN2K}, an all-electron electronic structure code based on the full-potential linearized augmented plane wave (FP-LAPW) method \cite{wien2k}. The exchange-correlation potential is constructed using generalized gradient approximation (GGA) in the parametrization scheme of Perdew, Burke, and Ernzerhof (PBE)~\cite{PBE}. The muffin tin radii for different atoms were chosen to be nearly equal, non-overlapping, but as large as possible thus ensuring minimization of interstitial region and accurate determination of the atomic magnetic moments. The plane wave cut-off parameters were decided by $R_{\rm mt}K_{\rm max}$=8, where $R_{\rm mt}$ and $K_{\rm max}$ are the radius of the smallest muffin tin sphere and the largest wave vector of the basis set, respectively. $G_{\rm max}$=14 a.u.$^{-1}$ is employed for the Fourier expansion of potential in the interstitial region. The maximal angular quantum number $l_{\rm max}$ for the expansion of potentials and wave functions in the muffin tins is set to 10. For Brillouin zone sampling a dense $35\times 35\times 35$ Monkhorst-Pack $\vec{k}$-mesh was adopted. As a starting point for each set of calculations, the lattice parameter was optimized through the total energy versus volume curve by fitting the Murnaghan's equation of state~\cite{Murnagan's}. To address the mechanical stability of compounds, we carried out phonon dispersion calculations based on density functional perturbation theory (DFPT) \cite{Baroni}, implemented by the plane-wave pseudo-potential package {\scriptsize QUANTUM-ESPRESSO} \cite{quantum espresso}. For the electronic (self-consistent field) part of the calculations, the same GGA-PBE exchange-correlation functional as in the above was employed although with Vanderbilt pseudo-potentials and plane-wave basis sets. To calculate the force constants, a $4\times 4\times 4$ irreducible $\vec{q}$ point grid was constructed for each compound at its optimized lattice constant. It is worth pointing out that the total energy and electronic structure calculations performed by {\scriptsize QUANTUM-ESPRESSO} (not shown), essentially reproduce and therefore corroborate the results obtained by {\scriptsize WIEN2K}. Self-consistency in all calculations is considered to be achieved when the total energy converges to better than $10^{-6} $ Ry/f.u.\ (Rydberg per formula unit).

\begin{figure}[H]
\centering
\includegraphics[width =0.45\textwidth]{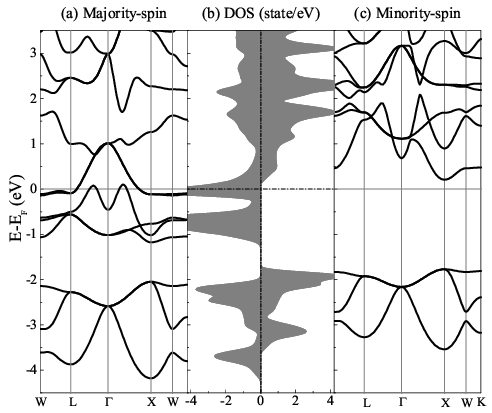}
\caption{Spin-resolved band structure and density of states of HH-VSbSr in $\beta$-phase at optimized lattice constant. The shaded regions represent the total density of states.} \label{Fig1}
\end{figure}

\section{prototype conventional Half-metal\label{sec:VSbSr}}
\subsection{Electronic structure of VSbSr}
Figure 1 shows the band structure of HH-VSbSr near the Fermi level for majority-spin (a), and minority-spin (c) at the optimized cubic lattice constant $a=7.24$\,{\AA}. The corresponding total density of states (DOS), is shown in panel (b) of the same figure. Evidently there is a semiconductor energy gap around the Fermi level in the minority-spin channel, while the majority-spin band structure exhibits a strong metallic character due to a large density of states at the Fermi level. The half-metallic gap or the minimum energy required for spin-flip excitation is $E_{\rm HM}=0.21$\,eV, which ideally results in 100\% spin-polarized conduction electrons, i.e.\ half-metallicity.

\begin{figure}[H]
\centering
\includegraphics[width =0.45\textwidth]{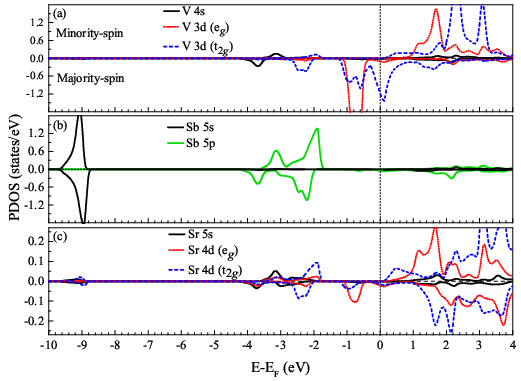}
\caption{(Color online) Spin-resolved site and orbital projected density of states of HH-VSbSr in the $\beta$-phase.} \label{Fig2}
\end{figure}

The calculated site and orbital projected density of state (PDOS) of HH-VSbSr at the optimized lattice constant is shown in Fig.\ 2. Clearly, there is a large spin splitting of the TM atom $d$ states across the Fermi level as can be seen in Fig.\ 2 (a). Because the electronegativity of middle $sp$ atom $X_{\rm Sb}=2.05$ is closer to that of the TM atom $X_{\rm V}=1.63$, as compared to the other neighbor $X_{\rm Sr}=0.95$, there is a stronger tendency for covalent hybridization between Sb and V. So, similar to the zinc-blende TM pnictides \cite{Liu}, the half-metallic gap arises primarily from the $p$-$d$ bonding and antibonding covalent interactions between the $sp$ atom $p$ and the TM atom $d$ states of $t_{\rm 2g}$ subspace, which is sometimes referred to as the `$p$-$d$ exchange' in the literature \cite{Galanakis13,Galanakis02}. It must be noted that the local tetrahedral crystal field splits the $d$ states of TM atom into three-fold degenerate $t_{\rm 2g}$ and two-fold degenerate $e_{\rm g}$ with the $t_{\rm 2g}$ subspace slightly lower in energy than the $e_{\rm g}$. The spin-resolved PDOS shown in Fig.\ 2 also indicates that, rather similar to the TM atom \cite{Liu}, the $s$ state of the $d^0$ metal atom hybridizes with its $d$ states. Compared to the TM atom, however, there is a much weaker covalent hybridization between the $d$ states of $d^0$ metal atom and the $p$ states of $sp$ atom because of large energy difference between the two as the $d$ states of $d^0$ metal atom are in excess of 4\,eV above the Fermi level. This is consistent with the large electronegativity  mismatch between the $d^0$ metal atom and the $sp$ atom. The interaction between the two is therefore of ionic character with a substantial transfer of valence $s$ electrons from Sr to Sb.

\begin{figure}[H]
\centering
\includegraphics[width =0.45\textwidth]{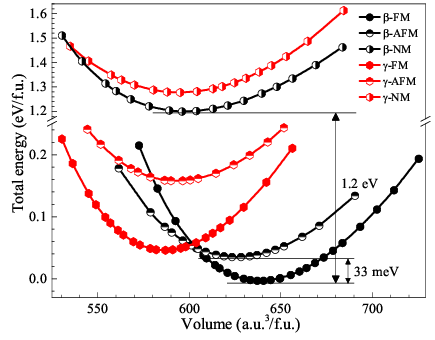}
\caption{(Color online) The total energy vs.\ volume curves for half-Heusler VSbSr in different competing phases and magnetic states. The minimum energy $\beta$-phase in ferromagnetic state, is taken as the zero of energy.} \label{Fig3}
\end{figure}

\subsection{Structural stability of VSbSr}
For the practical purposes of experimental realization and potential applications, the questions of mechanical and thermodynamic stability of the compounds must be addressed. In Fig.\ 3 the stability of HH-VSbSr in $\beta$-phase is checked against the rival $\gamma$-phase by calculating the total energy vs.\ volume curves in three different magnetic states, i.e. ferromagnetic (FM), [001] antiferromagnetic (AFM), and non-magnetic (NM) state. Similar to the case of lithium-based half-Heusler alloys \cite{Damewood}, here too the $\beta$-phase is found to be the minimum  energy configuration with a ferromagnetic alignment of the TM atoms. The calculations presented in Sec.\ \ref{sec:MnPK} show similar results for the other $d^0$-$d$ prototype HH-MnPK. It is therefore a matter of considerable interest to find that unlike the half-Heusler lithiated TM pnictides that are ordinary ferromagnets \cite{Damewood}, the $d^0$-metal-based TM pnictides such as HH-VSbSr do display ferromagnetic {\em half-metallicity} at optimized lattice constants. The ferromagnetic state is found to be 33 meV/f.u.\ more stable than the AFM state as shown in Fig.\ 3. To study the AFM state, a tetragonal unit cell consisting of two formula units was constructed to allow for antiparallel alignment of the TM atoms along the [001] direction. The HH $\alpha$-phase with the $d^0$ metal atom at the unique middle site is not shown in Fig.\ 3 because it is found to be electronvolts instable with respect to the minimum energy $\beta$-phase. The fact that the most electronegative atom sits in between and binds the two electropositive metallic atoms, stabilizes the HH structure in the $\beta$-phase rather similar to the lithium-based half-Heusler alloys.

\begin{figure}[H]
\centering
\includegraphics[width =0.45\textwidth]{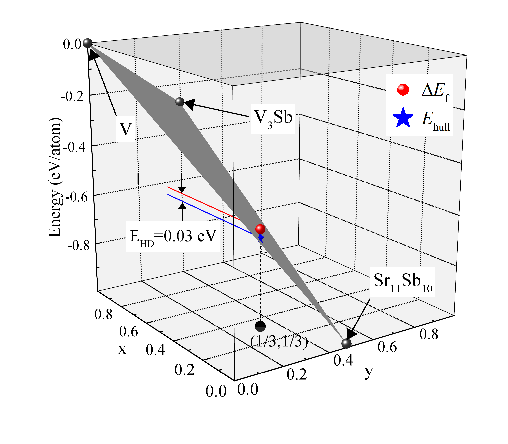}
\caption{(Color online) The formation energy of V$_x$Sb$_y$Sr$_{1-x-y}$ as a function of chemical composition ($x$,$y$). The dark shaded area shows the convex hull that is determined by the plane passing through the three phase points corresponding to the stable phases of the constituent elements that VSbSr can decompose into, i.e.\ Sr$_{11}$Sb$_{10}$, V$_{3}$Sb, and V, as extracted from the Open Quantum Materials Database (see text for more details).} \label{Fig5}
\end{figure}

To check the stability of material against phase separation, the calculation of formation energy is always the first step. For a ternary half-Heusler compound ABC, the formation energy per formula unit (at zero temperature and pressure conditions) is given by
\begin{equation}
\Delta{E_{\rm f}}=E_{\rm ABC}-(E_{\rm A}+E_{\rm B}+E_{\rm C}),
\label{Eq.1}
\end{equation}
where $E_{\rm ABC}$ and $E_{i}$ are the total energy per formula unit of the HH compound ABC and that of the elemental constituent \textit{i}, respectively, obtained by first principles DFT calculations. A negative value of the formation energy indicates that at zero temperature and pressure conditions a compound is more stable than its constituent elements. This must, however, be regarded as a {\em necessary but not sufficient} condition for ground state thermodynamic stability as it does not guarantee the stability of a compound against decomposition to all possible combinations of the constituents. An appropriate criterion for thermodynamic stability is the convex hull construction, which compares the formation energy of the structure in question to those of other stable phases formed by the constituent elements at that composition \cite{Ma17, Sanvito}.
Every phase on the convex hull has a formation energy lower than or equal to any other phase or linear combination of phases in the chemical space at that composition. Clearly, for calculating the convex hull of a ternary system A-B-C, one needs to investigate all possible compounds that can be formed from elements A, B, and C. Inevitably, one has to limit the search space of considered phases to a DFT database. We therefore used the Open Quantum Materials Database (OQMD) \cite{OQMD}, which is an extensive database consisting of DFT-predicted crystallographic parameters and formation energies for all binary and ternary phases experimentally observed and reported in the International Crystal Structure Database (ICSD) \cite{ICSD}, plus more than 550,000 hypothetical compounds based on common structural prototypes.
The distance from the convex hull $E_{\rm HD}$ for a phase with formation energy $\Delta{E_{\rm f}}$ is given by
\begin{equation}
E_{\rm HD}=\Delta{E_{\rm f}}-E_{\rm hull},
\label{Eq.2}
\end{equation}
where $E_{\rm hull}$ is the energy on the convex hull at the composition of the phase in question. The distance of a phase from the convex hull, $E_{\rm HD}$, can be considered as an indicator of the likelihood of its synthesis in experiments, which are of course at finite temperature and pressure conditions (unlike DFT calculations). It has recently been reported that a hull distance of less than $\sim$0.1 eV/atom is the approximate threshold for successful experimental  synthesis of  half-Heusler alloys with some notable exceptions \cite{Ma17}.

Practically, the phase stability of a ternary compound A$_{x}$B$_{y}$C$_{1-x-y}$, where $0\leq x\leq 1$ and $0\leq y\leq 1-x$, is studied in a three-dimensional $xyE$-space. In this space every chemical phase is represented by a phase point ($x, y, E$), where $E$ denotes the corresponding formation energy per atom. In this framework, the convex hull can be formed by the planes lowest in energy that pass through any three stable phase points. If the formation energies of the relevant stable binaries (such as A$_{x}$B$_{1-x}$, A$_{x}$C$_{1-x}$, and B$_{x}$C$_{1-x}$ with $0\leq x\leq 1$) are available from a DFT database, then one can readily construct the convex hull diagram relevant for the associated A$_{x}$B$_{y}$C$_{1-x-y}$ ternary compounds. The information obtained from OQMD reveals that the most preferred decomposition path of the ternary compound VSbSr, also considering stoichiometry constrains, is given by
\begin{equation}
\rm VSbSr \longrightarrow \dfrac{1}{11} Sr_{11}Sb_{10} + \dfrac{1}{11} V_{3}Sb + \dfrac{8}{11} V
\label{Eq.3}
\end{equation}
where, Sr$_{11}$Sb$_{10}$, V$_{3}$Sb, and V are the relevant stable phases, which can be mapped onto three phase points ({\small 0}, $\frac{10}{21}$, {\small $-$0.984\,eV/atom}), ($\frac{3}{4}$, $\frac{1}{4}$, {\small $-$0.213\,eV/atom}), and ({\small 1}, {\small 0}, {\small 0}) in the $xyE$-space, respectively, as shown in Fig.~\ref{Fig5}. The convex hull is given by the plane passing through these stable phase points and the energy at the composition ($x=\frac{1}{3}$, $y=\frac{1}{3}$) in this plane specifies the energy of the convex hull $E_{\rm hull}$. The red ball and the blue star in Fig.~\ref{Fig5} represent the formation energy $\Delta{E_{\rm f}}$ = $-$0.62\,eV/atom of HH-VSbSr and the energy of the convex hull $E_{\rm hull}$ = $-$0.65 eV/atom at the same composition, respectively. Thus, by Eq.\ \ref{Eq.2}, the distance from convex hull becomes $E_{\rm HD}$ = 0.03 eV/atom. It is clear that the formation energy of HH-VSbSr is very close to the convex hull with a hull distance much lower than the empirical threshold ($\sim$0.1 eV/atom) for successful empirical synthesis. This study therefore indicates that HH-VSbSr may be realizable by equilibrium methods such as chemical synthesis. Clearly, HH-VSbSr is a good candidate for further experimental investigation. To highlight the importance of the role played by the $d^{0}$ atom, it must be pointed out that the distance from convex hull of the half-metallic TM pnictides in ZB phase is in general much larger, which explains why they have only been grown epitaxially in ultra-thin film form, not exceeding ten atomic layers \cite{A05}. In this case too, the distance of ZB-VSb from the convex hull is at least 0.14\,eV/atom \cite{MP}. Hence, although `$p$-$d$ exchange' is the common source of half-metallicity in both HH-VSbSr and ZB-VSb, but HH structure has a substantially lower hull distance than the ZB structure, rendering HH-VSbSr much more favorable to experimental realization.

\begin{figure}[H]
\centering
\includegraphics[width =0.45\textwidth]{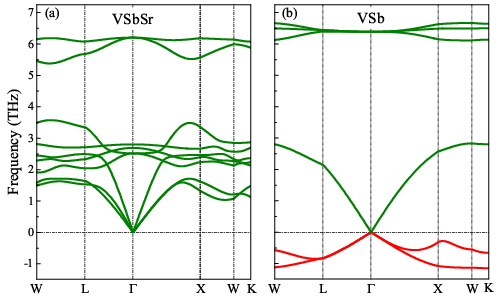}
\caption{(Color online) The phonon band structure of HH-VSbSr (a), and that of ZB-VSb (b), at the corresponding optimized lattice constants.} \label{Fig4}
\end{figure}

Furthermore, the absence of negative frequency modes in the phonon band structure, reveals that the structure of HH-VSbSr is dynamically stable with respect to perturbations of all wave vectors.  Figure \ref{Fig4} shows the phonon band structure of HH-VSbSr (a), and that of the corresponding TM pnictide VSb in the  half-metallic zinc-blende phase (b). While ZB-VSb displays negative frequency transverse acoustic modes, which is a clear indication of mechanical instability with respect to shear strain, HH-VSbSr is devoid of such instabilities. Therefore, HH-VSbSr has all the desirable properties of ZB-VSb with an added bonus, i.e.\ {\em true metastability}, which means that HH-VSbSr lies in an energy minimum of the configuration space. So in the case of likely experimental realization indicated by the small hull distance, the $d^0$-$d$ half-Heusler alloy VSbSr tends to maintain its structure by true metastability.

\subsection{Magnetic properties of VSbSr}
The Slater-Pauling (SP) rule for the total magnetic moment (in units of Bohr magneton) $M_{\rm tot}=Z_{\rm tot} - 2N_{\rm min}$, where $Z_{\rm tot}$ is the total number of valence electrons per formula unit, is obtained by counting the number of valence electrons per formula unit in the minority-spin, $N_{\rm min}=4$, which accounts for the occupied three $p$ states and the single $s$ state of $sp$ atom deep in energy as can be seen in Fig.\ 1 (c) and Fig.\ 2 (b), respectively. For the $d^0$-$d$ HH alloys, therefore, the SP rule takes the form
\begin{equation}
M_{\rm tot}=Z_{\rm tot} - 8.
\end{equation}
This is the same SP rule as that found for the TM pnictides \cite{Galanakis13,Galanakis02}.
The total magnetic moment per formula unit of VSbSr is found to be $M_{\rm tot}= 4.00$, which is consistent with the above SP rule. The atomic magnetic moments are $M_{\rm V}= 3.63$, $M_{\rm Sb}= -0.11$, and $M_{\rm Sr}= 0.08$, which together with the magnetic moment in the interstitial region $M_{\rm int}=0.40$,  sum up to $M_{\rm tot}=4.00$. Consistent with the large $d$-band spin splitting of the V atoms, shown in Fig.\ 2 (a), the bulk of magnetic moment is carried by the TM atom. The ferromagnetic coupling of V atoms gives the magnetic ground state. As it is evident from Fig.\ 3, the difference in total energy between non-magnetic and ferromagnetic state is $\Delta E_{\rm NM-FM}= 1.20$\,eV/f.u. On inserting this into the Heisenberg spin-Hamiltonian \cite{Sivadas15}, the nearest-neighbor exchange coupling is found to be $J_{\rm V}= \Delta E_{\rm NM-FM}/(12 |M_{\rm V}|^2)=7.59$~\,meV, where 12 in the denominator accounts for the number of fcc nearest neighbors. By Monte Carlo simulation of nearest-neighbor Ising model on the fcc sublattice of V atoms, and finite-size scaling of the fourth-order cumulant \cite{Binder}, the Curie temperature of HH-VSbSr is found to be $T_{\rm c}= 9.79\,J_{\rm V}/k_{\rm B} = 862$\,K, where $k_{\rm B}$ is the Boltzmann constant. The high Curie temperature thus obtained, is indicative of the magnetic stability of the compound for potential room temperature applications. It is worth mentioning that we find the same procedure to give $T_{\rm c}=723$\,K for the first known half-metal, MnNiSb, which is within a percentage error of the corresponding experimental value $730$\,K.

\begin{figure}[H]
\centering
{\includegraphics[width =0.45\textwidth]{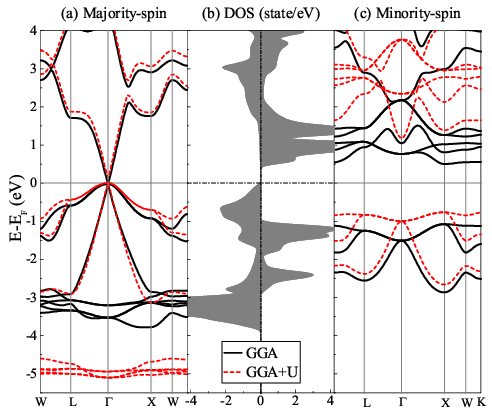}\hspace*{-0.07cm}}
\caption{(Color online) Spin-resolved band structure and total density of states of HH-MnPK in $\beta$-phase at optimized lattice constant. The dashed red lines represent the band structure obtained from GGA+U exchange-correlation approximation.} \label{Fig6}
\end{figure}

\section{Prototype Dirac half-metal \label{sec:MnPK}}
\subsection{Band structure of HH-MnPK}

The zero-band-gap behavior with either linear or parabolic energy dispersion, has attracted considerable attention owing to both a fundamental interest in a new state of matter and its possible application in new spintronic devices~\cite{Nat1, Nat2}. Figure 6 shows the band structure and total DOS of HH-MnPK in $\beta$-phase for majority- and minority-spin at the optimized lattice constant $a=6.73$\,{\AA}. Evidently, a wide half-metallic gap $E_{\rm HM}=0.51$\,eV characterizes the minority-spin band structure. In the majority-spin channel, however, the conduction and the valence bands touch directly at the $\Gamma$-point and the Fermi level. The linear and steep dispersions of the conduction band and one of the valence bands in the majority-spin channel, indicates that both carriers (electrons and holes) have vanishingly small band-mass and high mobilities, in addition to both being 100\% spin-polarized. However, because there are three degenerate valence bands touching the conduction band at the $\Gamma$-point and the Fermi level, two of which have ordinary parabolic dispersions, there will be both massive and massless holes in majority-spin channel at finite temperatures. The Fermi velocity of Dirac fermions is found to be $v_{\rm F}= 7.10\times 10^5$\,m/s that is 71\% of that in graphene \cite{Graphene}. HH-MnPK by far shows the {\em highest Fermi velocity} compared to the 2D and quasi-2D DHM materials theoretically proposed so far \cite{Pardo09,Zhang15,Cai15,Wei16,He16}. To identify the origin of the Dirac point degeneracy, it is found that a 2\% symmetry-conserving hydrostatic lattice contraction opens a small gap, which shows that the Dirac point degeneracy is accidental and not related to crystal symmetries \cite{Pickett}. In order to check the stability of band structure with respect to static correlations, the GGA+U exchange-correlation approximation is employed with the on-site coulomb energy $U=4.0$\,eV for the Mn atom. As shown in Fig.\ 6, the band structure so obtained (dashed red lines) is essentially the same as the GGA-PBE band structure at and near the Fermi level. Furthermore, the inclusion of spin-orbit coupling (SOC) is found to have very little effect on the hybrid P $3s$ and $3p$ bands at and near the Fermi level, because P is a small atom.

\begin{figure}[H]
\centering
\includegraphics[width =0.45\textwidth]{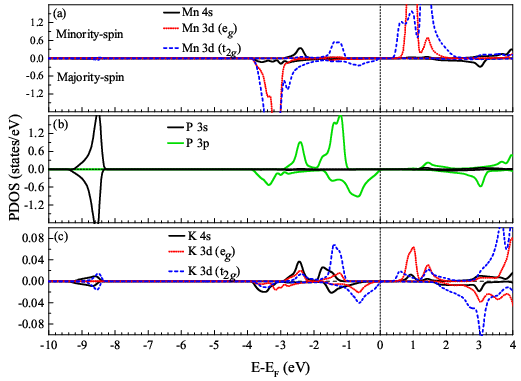}
\caption{(Color online) Spin-resolved site and orbital projected density of states of HH-MnPK in the $\beta$-phase.} \label{Fig9}
\end{figure}

To ascertain the character of bands, the spin-resolved PDOS of MnPK is shown in Fig.\ \ref{Fig9}. The K atom tends to give up its single valence electron to its highly electronegative P neighbor with very little covalent interaction. It must also be noted that $4s$ state of Mn mixes with its $3d$ states, thus forming six hybrid active states. Starting with the minority-spin bands, and apart from the single P $3s$ state deep in energy, there are three valence bands of primarily P $3p$ character in the range of energies $-3$ to $-1$\,eV below the Fermi level. These bands originate from the covalent bonding interaction between the three P $3p$ and the three Mn $3d$ states of $t_{\rm 2g}$ subspace. Delineating the half-metallic gap and in the range of energies $0.5$ to $1$\,eV above the Fermi level, there are three narrow bands of which two originate from the non-bonding $e_{\rm g}$ subspace of Mn $3d$ states, and the other is primarily of Mn $4s$ character particularly near the $\Gamma$ point as indicated by the band character (fat band) plots. The next three bands in the range 1 to 4\,eV stem from the corresponding antibonding states of the above-mentioned covalent bonding states. These bands are of Mn $t_{\rm 2g}$ character with an admix of P $3p$ of course. Similar features can also be found in the majority-spin channel, however, with a big distinction. While the minority-spin bands across the Fermi level are in the normal $p$-$d$ order (in ascending order of energy), which is rather similar to the conventional half-metals such as VSbSr, in the majority-spin channel the Mn $3d$ bands are inverted and pushed down below the valence P $3p$ bands by the huge $d$-band exchange splitting. The linear valence and conduction band are a mixture of P $3p$, P $3s$, and Mn $4s$ state near the $\Gamma$-point, while the parabolic valence bands are mainly of P $3p$ character. This explains why the inclusion of Hubbard $U$ had essentially no effect on the behavior of bands at and near the Fermi level shown in Fig.\ 6, although it resulted in an enhancement of $d$-band spin splitting by pushing the Mn $3d$ states further away from the Fermi level.

\begin{figure}[H]
\centering
\includegraphics[width =0.45\textwidth]{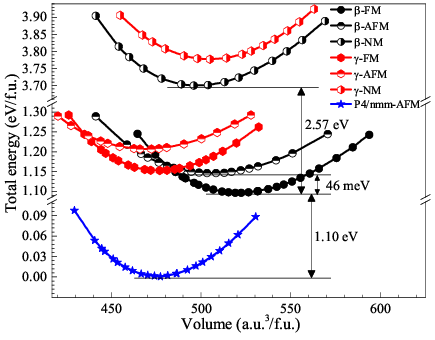}
\caption{(Color online) The total energy vs.\ volume curves for MnPK in different half-Heusler phases and magnetic states are compared with the ground state tetragonal phase (space group P4/nmm) in the antiferromagnetic state. The ground state minimum is taken as the zero of energy.} \label{Fig7}
\end{figure}

\subsection{Structural properties of HH-MnPK}
\begin{figure}[H]
\centering
\includegraphics[width =0.45\textwidth]{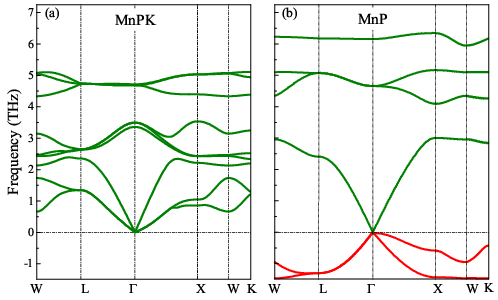}
\caption{(Color online) The phonon band structure of HH-MnPK (a) and that of ZB-MnP (b), at the corresponding optimized lattice constants.} \label{Fig10}
\end{figure}

As in the case of VSbSr, the total energy is calculated as a function of the volume for HH-MnPK in two different phases ($\beta$ and $\gamma$) and in three different magnetic states (FM, AFM, and NM), the results of which are summarized in Fig.\ \ref{Fig7}. Evidently, among the half-Heusler phases, $\beta$-phase is the most stable with a ferromagnetic alignment of the Mn atoms. It has, however, come to our notice that MnPK has been produced by chemical synthesis in the tetragonal phase \cite{MnPK}, i.e.\ space group P4/nmm (No.\ 129), and subsequent neutron scattering studies have found its magnetic structure to be AFM \cite{MnPK-GS}. We therefore treated this tetragonal AFM phase as the thermodynamic ground state of the compound, with which the half-Heusler FM phase is compared in Fig.\ 8. The distance from the convex hull, which in this case is simply the excess or metastable energy of HH-MnPK with respect to the ground state tetragonal phase, is one criterion for successful experimental synthesis.  The hull distance is found to be $E_{\rm HD} = 0.37$\,eV/atom, which is readily deducible from Fig.\ 8. Although the actual metastability of HH-MnPK is larger than the empirical criterion for the successful equilibrium synthesis of the half-Heusler alloys ($\lesssim 0.1$\,eV/atom) \cite{Ma17}, this does not rule out the possibility of synthesis by non-equilibrium techniques such as epitaxy. In fact compounds with relatively large distances from the convex hull, such as ZB-CrAs with a hull distance of 0.45\,eV/atom \cite{Xie}, have been synthesized epitaxially in the ultra-thin film form \cite{Akinaga}. This is despite the fact that ZB-CrAs is not even a true metastable phase. To show that HH-MnPK is indeed a true metastable phase, the phonon band structure of HH-MnPK is compared with that of the corresponding TM pnictide ZB-MnP in Fig.~\ref{Fig10}. Unlike ZB-MnP, HH-MnPK is devoid of imaginary frequencies. HH-MnPK is therefore dynamically stable. So, even though HH-MnPK cannot be synthesized by equilibrium means, non-equilibrium techniques such as epitaxy may work given that HH-MnPK is a true metastable phase. This with a negative formation energy $\Delta E_{\rm f} = -0.56$\,eV/atom,  simple stoichiometric 3D structure, and high Curie temperature (discussed next), make HH-MnPK a suitable DHM candidate for epitaxial growth and potential room temperature applications.

\subsection{Magnetic and other important attributes of HH-MnPK}

Because all the bands through valence $p$ are completely occupied in both spin channels, the considerations that lead to the Slater-Pauling rule, Eq.\ (4), remain valid. Consistent with the SP rule, the total magnetic moment is found to be $M_{\rm tot}= 5.00$, with the atomic distribution $M_{\rm Mn}= 4.65$, $M_{\rm P}= 0.02$, $M_{\rm K}= 0.04$, and $M_{\rm int}= 0.29$. HH-MnPK is therefore a {\em maximal moment} $d$-shell ferromagnet. The magnetic moment of K atom turns out to be negligible as it tends to give up its single valence electron to its highly electronegative P neighbor. This enhanced ionic character stabilizes the HH structure in $\beta$-phase rather similar to the case of HH-MnPLi \cite{Damewood}.
The Curie temperature of HH-MnPK is obtained using the same procedure employed for VSbSr.
The exchange coupling between the nearest neighbor Mn atoms $J_{\rm Mn} = \Delta E_{\rm NM-FM}/(12 |M_{\rm Mn}|^2)=9.90$\,meV, is found from the energy difference between non-magnetic and ferromagnetic state given in Fig.\ 8, which then leads to a Curie temperature $T_{\rm c}=9.79J_{\rm Mn}/k_{\rm B} = 1120$\,K. The very high Curie temperature together with the wide half-metallic gap, indicate that the HM property in MnPK is robust with respect to the collapse of spin-polarization with the temperature \cite{Lezaic,Aldous}.

A significant feature of the majority-spin bands in HH-MnPK is the $d$-$p$ band-inversion, while the minority-spin bands are in the normal $p$-$d$ order across the Fermi level. This band inversion is driven by neither SOC nor lattice strain. It may, however, be related to the maximal $d$-shell magnetic moment through the huge exchange splitting of the Mn $d$ bands, as can be noted in Fig.~\ref{Fig9} (a). The exchange splitting is roughly given by $I\cdot M_{\rm Mn}\sim 4.2$\,eV, where $I\sim 0.9$\,eV is the exchange integral of TM atoms.
This $d$-$p$ band-inversion is a clear indication of a topologically non-trivial electronic structure that leads to a quantum anomalous Hall state \cite{Liu08}. As also pointed out by Wang \cite{Wang16}, the DHM is at the critical point of transition to a magnetic topological or Chern insulator through the application of lattice strains that open a band gap \cite{Xiao10,Nat1,Nat2,Casper12}, and thus result in a quantum anomalous Hall effect that is characterized by 100\% spin-polarized, massless, and dissipationless surface states \cite{Haldane88}. DHM materials such as HH-MnPK are therefore of vital interest for ultra-fast and power-efficient spintronics of the future.

\begin{figure}[H]
\centering
\includegraphics[width =0.45\textwidth]{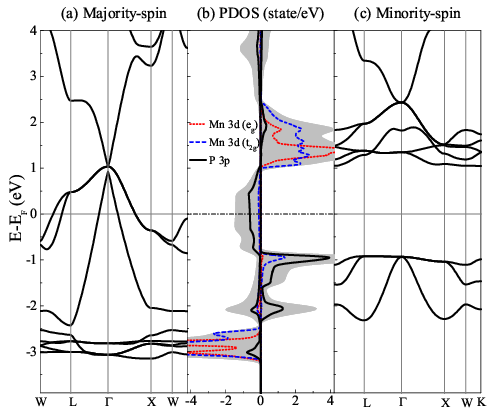}
\caption{(Color online) Spin-resolved band structure and partial density of states of ZB-MnP at the {\em expanded} lattice parameter $a=6.73$\,{\AA}. The shaded regions represent the total density of states.} \label{Fig8}
\end{figure}

To clarify the significance of the role played by the $d^{0}$ atom K, the band structure of ZB-MnP at the lattice constant of HH-MnPK, $a=6.73$\,{\AA}, is shown in Fig.~\ref{Fig8}, which is essentially the band structure of HH-MnPK less the K atom.
By comparing Fig.\ 10 with the band structure of HH-MnPK shown in Fig.\ 6, one can readily notice that the two have much in common. In particular, in both cases the covalent hybridization of P $p$ and Mn $t_{\rm 2g}$  states, i.e.\ the `$p$-$d$ exchange', is mainly responsible for the half-metallic gap in the minority-spin channel. Furthermore, in Fig.\ 10 there is a Dirac point situation at the $\Gamma$-point in the majority-spin channel, rather similar to that in Fig.\ 6, albeit an electron-volt above the Fermi level. It must also be noted that ZB-MnP at its optimized lattice constant, $a=5.31$\,\AA, is an ordinary metallic ferromagnetic \cite{GM03,Conti}. So, although it has no role in covalent interactions, the addition of $d^0$ atom K has three important consequences: (i) the expansion of ZB-MnP to the extent that results in a wide half-metallic gap as shown in Fig.\ 10 (c), (ii) addition of an extra valence electron to the compound that sets the Fermi level right on the Dirac point as seen in Fig.\ 6 (a), and (iii) producing true metastability in the otherwise unstable structure of ZB-MnP as shown in Fig.\ 9.

\section{\label{sec:Summary}Conclusions}
It was shown systematically by first principles density functional calculations that the $d^0$-$d$ half-Heusler alloys formed by $3d$ transition metals and $d^0$ alkali or alkaline-earth metals defined by the valence electronic configuration $ns^{1,2}$, $(n-1)d^0$ can produce conventional as well as Dirac half-metallicity, which is reported for the first time in a three-dimensional prototype HH-MnPK. Together with the predicted mechanical stability and high Curie temperature, this finding could pave the way for massless and dissipationless spintronics of the future. HH-MnPK can also serve as a prototype for further theoretical investigation into the topological states of 3D matter as there are clear indications of an inverted band order across the Fermi level in one of the spin channels. Furthermore, using HH-VSbSr as a prototype conventional half-metal, it was shown that the introduction of $d^0$ metal atom Sr reduces the distance from the convex hull and produces true metastability in the otherwise unstable structure of transition metal pnictides in zinc-blende phase without altering the `$p$-$d$ exchange' that is mainly responsible for their half-metallicity, thus making their realization and potential application in spintronics feasible. We hope that the above theoretical results will indeed inspire and facilitate the epitaxial growth of $d^0$-$d$ half-Heusler alloys on zinc-blende or half-Heusler substrates of compatible lattice parameter. At this stage, therefore, a thorough experimental investigation of $d^0$-$d$ HH alloys is warranted.

\section*{ACKNOWLEDGEMENTS}

The authors are indebted to Prof.\ Warren E. Pickett for a number of illuminating communications.


\end{document}